\documentclass[12pt]{article}
\usepackage{graphics,cite}

\def \gsim
{\raisebox{-3pt}{$\>\stackrel{>}{\scriptstyle\sim}\>$}}

\begin{document}
\begin{titlepage}
\begin{flushright}
MPP-2003-60
\end{flushright}
\par \vspace{10mm}
\begin{center}
{\Large \bf
Soft-Gluon Resummation\\[1ex]
for Heavy Quark Production\\[2ex]
in Charged-Current Deep Inelastic Scattering}
\end{center}
\par \vspace{2mm}
\begin{center}
{\large \bf G. Corcella $^1$ and A.D. Mitov $^2$}\\
\vspace{2mm}
{$^1$ Max-Planck-Institut f\"ur Physik, Werner-Heisenberg-Institut,}\\
{F\"ohringer Ring 6, D-80805 M\"unchen, Germany \@.}\\
\vspace{2mm}
{$^2$ Department of Physics and Astronomy, University of Rochester,\\
Rochester, NY 14627, U.\ S.\ A.}\\
\end{center}
\par \vspace{2mm}
\begin{center}
{\large \bf Abstract}
\end{center}
\begin{quote}
  \pretolerance 10000
We study soft-gluon radiation for heavy quark production in
charged-current Deep Inelastic Scattering processes. 
We resum large-$x$ contributions to the 
$\overline{\mathrm{MS}}$ coefficient function
to next-to-leading logarithmic accuracy and present 
results for charm quark structure functions
for several values of the ratio $m_c/Q$.
The effect of soft resummation turns out to
be visible, especially at small $Q^2$, and the results exhibit
very little dependence on factorization and
renormalization scales.
The impact of our calculation on structure function measurements at 
NuTeV and HERA experiments is discussed.
\end{quote}
\end{titlepage}

\section{ Introduction}
Lepton-hadron Deep Inelastic Scattering (DIS) is one of the most interesting 
processes to investigate Quantum Chromodynamics,
from both theoretical and experimental point of view.
In fact, DIS allows one to 
measure proton structure functions, and hence 
investigate the internal structure of hadrons.
For the sake of performing accurate measurements of observables
in DIS processes, precise QCD calculations are necessary.

In this paper we study heavy quark production in DIS, 
in particular charm quark production in charged-current (CC) events.
Measurements of the charm quark structure functions 
in the CC regime are in fact important
in order to probe the density of strange quarks
and gluons in the proton.

Charged-current Deep Inelastic Scattering processes 
$\nu_\mu N\to \mu X$ are currently
under investigation at the NuTeV experiment at Fermilab,
where neutrino and antineutrino 
beams collide on an fixed iron target (see, e.g., 
Ref.~\cite{nutev} for the updated results). In particular, production
of two oppositely-charged muons at NuTeV \cite{nutev1} is mainly 
associated with a CC event with a charm quark in the final state.

The first generation of H1 and ZEUS experiments 
on electron-proton DIS at the HERA collider 
at DESY (HERA I) did detect charged-current events $ep\to \nu_eX$
\cite{h1,zeus}, 
but it did not have sufficiently-high statistics
to reconstruct heavy quarks in the CC regime. However, 
such measurements are foreseen in the current improved-luminosity
run (HERA II) and at the possible future
generation of HERA III experiments \cite{hera3}, 
which should also be able to
investigate the proton structure functions at low $Q^2$ values.

Structure functions are given by convolutions 
of short-distance, partonic coefficient functions, $C(x,\mu_F^2)$,
usually given in the $\overline{\mathrm{MS}}$ factorization scheme, with
long-distance parton 
distribution functions $f(x,\mu_F^2)$, whose dependence on the 
factorization scale $\mu_F$ is ruled by the 
Dokshitzer--Gribov--Lipatov--Altarelli--Parisi (DGLAP) evolution equations
\cite{ap,dgl}.
The next-to-leading order (NLO)  
$\overline{\mathrm{MS}}$ coefficient functions for charm production
in CC DIS present terms which get arbitrarily large once the
incoming quark energy fraction $z$ in $qW^*\to c(g)$ approaches unity.
This corresponds to soft-gluon radiation. 
In order to reliably predict structure functions at large $z$, such terms
need to be resummed to all orders in the strong coupling constant $\alpha_S$ 
(threshold resummation).
It is the purpose of this paper to perform the resummation of
the $\overline{\mathrm{MS}}$ coefficient function and investigate the
phenomenological implications of the resummation
on charm quark structure functions at NuTeV and HERA.

Soft resummation for the coefficient function in DIS was
first considered in \cite{ct,ster,cmw} and later 
implemented in \cite{vogt}, but in the approximation in 
which all participating quarks are massless. 
Soft-gluon resummation for heavy quark production
in DIS processes has been investigated in \cite{laenen,nkoy}, where
the authors have considered neutral current interactions.
Phenomenological studies at NLO on charm quark production 
at NuTeV and HERA
have been performed in \cite{kmo} and \cite{cagre} respectively.

The plan of our paper is the following.
In Section 2 we review the NLO result for heavy quark production in
CC DIS and discuss the behaviour of the coefficient function for
soft-gluon radiation.
In Section 3 we present results for the soft-resummed
$\overline{\mathrm{MS}}$ coefficient function. In particular, we shall
compare the results on CC DIS with 
soft resummation in top quark decay \cite{cm,ccm} and we shall 
comment on the differences with respect to the previous work
on light quark production in DIS \cite{ct,cmw,vogt}.
In Section 4 we present our 
predictions on charm quark structure functions for several values
of $Q^2$ which are relevant for NuTeV and HERA phenomenology.
In Section 5 we shall summarize the main results of our work
and discuss its possible future extensions.

\section{Coefficient functions and structure functions at NLO}

In this section we would like to review the main results on 
CC DIS at NLO in the strong coupling constant $\alpha_S$. 
The Born parton-level process in CC DIS reads:
\begin{equation}
q_1(p_1)W^*(q)\to q_2(p_2). 
\label{born}
\end{equation}
Process (\ref{born}) receives ${\cal O}(\alpha_S)$ corrections from 
quark-scattering 
\begin{equation}
q_1(p_1)W^*(q)\to q_2(p_2) \left( g(p_g) \right)
\label{qs}
\end{equation}
and gluon-fusion process
\begin{equation}
g(p_g)W^*(q)\to \bar q_1(p_1)q_2(p_2).
\label{gf}
\end{equation}

Hereafter, we shall assume that $q_1$ is light and $q_2$ is heavy,
i.e. $p_1^2=0$ and $p_2^2=m^2$. 
We also define $Q^2=-q^2$ and the Bjorken $x$ as
\begin{equation}
x={{Q^2}\over{2P\cdot q}},
\end{equation}
where $P$ is momentum of the nucleon upon which the incoming
lepton scatters, i.e.
a proton for $ep$ interaction, a proton or a neutron for $\nu N$ processes.

In the following, in order to appreciate quark mass effects,
we shall be mainly interested in values of
$Q^2$ with respect to which $m^2$ is not 
negligible. Nevertheless, we shall present results for both small and large
values of $m^2/Q^2$ and comment on the
transition between the massive case and the massless approximation.

At small values of $Q^2$ one should also take into account the
effect of the target mass $M$, typically of the order of 1 GeV.
This leads to the introduction of the Nachtmann variable $\eta$, 
which generalizes the Bjorken $x$ to the case of non-negligible target mass.
It satisfies the following relation:
\begin{equation}
{1\over\eta} ={1\over 2x} +\sqrt{{1\over 4x^2}+{M^2\over
Q^2}}.\label{eta}
\end{equation}
The coefficient function of the quark-scattering process (\ref{qs}) is
usually expressed in terms of the variable
\begin{equation}
z={{Q^2+m^2}\over{2p_1\cdot q}},
\label{zeta}
\end{equation}
with $0\leq z\leq 1$.
Here we also introduce the energy fraction $\xi$ of $q_1$ with respect to
the initial-state proton\footnote{$\xi$ can be defined, e.g., in a frame 
where the nucleon and the quark are collinear \cite{aot}. Given 
$P=\left(P^{(0)},0,0,P^{(3)}\right)$ and 
$p_1=\left(p_1^{(0)},0,0,p_1^{(3)}\right)$, one has:
$\xi=\left(p_1^{(0)}+p_1^{(3)}\right)/\left( P^{(0)}+P^{(3)} \right)$.}
and the quantity
\begin{equation}
\chi={\eta\over\lambda},
\label{chi}
\end{equation}
where we have defined
\begin{equation}
\lambda={{Q^2}\over{Q^2+m^2}}.
\label{lamb}
\end{equation}
In the approximation of vanishing target and quark masses, $\chi$
coincides with the Bjorken $x$, i.e. $\chi\to x$.
Moreover, the relation $z=\chi/\xi$ holds and
$x$ and $\chi$ are related via \cite{aot}:
\begin{equation}
x={{\lambda \chi}\over{1-M^2\lambda^2\chi^2/Q^2}}.
\end{equation}
The Bjorken $x$, which in the massless approximation varies between 0 and
1, is now constrained in the range:
\begin{equation}
0< x\leq {{\lambda}\over{1-M^2\lambda^2/Q^2}}.
\end{equation}

The differential distributions $d\hat\sigma/dz$
of the parton-level processes in Eqs.~(\ref{qs}) and
(\ref{gf}) are collinear divergent and
their explicit expressions can be found in \cite{got}.
Such collinear singularities are usually regularized in dimensional 
regularization and subtracted in the $\overline{\mathrm{MS}}$ 
factorization scheme. This procedure yields the $\overline{\mathrm{MS}}$
coefficient functions \cite{gkr}.

Following the notation of \cite{kr}, and referring to neutrino
scattering on a fixed target, the inclusive 
cross section for $\nu(\bar\nu)$ scattering in the CC regime 
can be expressed as a linear combination of
three structure functions $F_1$, $F_2$ and $F_3$:
\begin{equation}
{d^2\sigma^{\nu(\bar\nu)}\over dxdy} = {G_F^2ME\over
\pi(1+Q^2/m_W^2)^2}\left\{y^2xF_1 +
\left[1-\left(1+{{Mx}\over{2E}}\right)y\right]F_2
\pm y\left(1-{y\over 2}\right)
x F_3\right\}. \label{sigma}
\end{equation}
In the above equation, $G_F$ is the Fermi constant, 
$E$ is the neutrino energy in the target rest 
frame and $y$ the inelasticity variable, given by:
\begin{equation}
y={{P\cdot q}\over{P\cdot k}},
\end{equation}
with $k$ being the momentum of the incoming lepton, the neutrino in
the case of Eq.~(\ref{sigma}).
In Eq.~(\ref{sigma}), we have ignored structure
functions $F_4$ and $F_5$ since their contribution to the
differential cross section is proportional to powers of the lepton 
masses and is therefore suppressed \cite{kr}.
Analogous expressions to Eq.~(\ref{sigma}) hold for positron (electron) 
scattering on a proton as well.

Following \cite{gkr,kr,ks}, one defines the `theoretical' structure functions
${\cal F}_i$ which are related to the $F_i$'s of Eq.~(\ref{sigma}) 
via the following relations:
\begin{eqnarray}
F_1 &=& {\cal F}_1,\label{calf1}\\
F_2 &=& {{2x}\over{\lambda\rho^2}} {\cal F}_2,\label{calf2}\\
F_3 &=& {2\over\rho} {\cal F}_3,
\label{calf3}
\end{eqnarray}
where we have defined:
\begin{equation}
\rho=\sqrt{1+\left( {{2Mx}\over Q} \right)^2}.
\end{equation} 

The introduction of the structure functions $\mathcal{F}_i$ is 
convenient as they can be straightforwardly expressed as
the convolution of parton distribution functions with
$\overline{\mathrm{MS}}$ coefficient functions:
\begin{equation}
\mathcal{F}_i(x,Q^2) = \int_\chi^1 {{d\xi}\over \xi}\left[
C_i^q(\xi,\mu^2,\mu_F^2,\lambda)q_1\left({\chi\over \xi},\mu_F^2\right) +
C_i^g(\xi,\mu^2,\mu_F^2,\lambda)g\left({\chi\over \xi},\mu_F^2\right)\right],
\label{conv}
\end{equation}
for $i=1,2,3$.
In Eq.~(\ref{conv}), $\mu$ and $\mu_F$ are the renormalization and
factorization scales;
$C_i^q$ and $C_i^g$ are the coefficient functions 
for quark-scattering (\ref{qs}) and
gluon fusion (\ref{gf}) processes; $q_1(\xi,\mu_F^2)$ and
$g(\xi,\mu_F^2)$ are the initial-state light quark and gluon parton 
distribution functions in (\ref{qs}) and (\ref{gf}).
We have chosen $x$ as the argument of ${\cal F}_i$ since later 
we shall present results for the structure
functions in terms of $x$.

The coefficient functions can be evaluated 
perturbatively and, to $\mathcal{O}(\alpha_S)$, they read:
\begin{equation}
C_i^q(z,\mu^2,\mu_F^2,\lambda) = \delta(1-z) +{\alpha_S(\mu^2)\over
2\pi} H_i^q(z,\mu_F^2,\lambda),
\label{cq}
\end{equation}
\begin{equation}
C_i^g(z,\mu^2,\mu_F^2,\lambda) = {\alpha_S(\mu^2)\over 2\pi}
H_i^g(z,\mu_F^2,\lambda). 
\label{cg}
\end{equation}
The expressions for the functions $H_{1,2,3}$ with the collinear
divergences subtracted in the $\overline{\mathrm{MS}}$ factorization
scheme can be found in \cite{gkr}. 
For consistency, in Eq.~(\ref{conv}) one will have to use 
$\overline{\mathrm{MS}}$ parton distribution functions as well.

As we are inclusive with respect to the final-state heavy quark
$q_2$, the quark-scattering coefficient functions are free from mass
logarithms $\ln(m^2/Q^2)$. 
However, the functions $H_i^q$ present terms which
behave like $\sim 1/(1-z)_+$ or $\sim [\ln(1-z)/(1-z)]_+$ 
and are therefore enhanced once the quark energy fraction $z$ approaches 
one \cite{gkr}.
The $z\to 1$ limit corresponds to soft-gluon emission.
In Mellin space, such a behaviour corresponds to contributions 
$\sim\ln N$ or $\sim\ln N^2$ in the limit $N\to\infty$, where the 
Mellin transform of a function $f(z)$ is defined as:
\begin{equation}
f_N = \int_0^1dz z^{N-1} f(z).
\end{equation}
The gluon-initiated coefficient functions $H_i^g$ in Eq.~(\ref{cg})
are however not enhanced in
the limit $z\to 1$, since the splitting $g\to q\bar q$ is not 
soft divergent.

We would like to make further comments on the coefficient functions at large 
$z$. Omitting terms that are not enhanced in the soft limit for
{\it any} value of the mass ratio $m^2/Q^2$, one has:
\begin{equation}
H^q_i(z\to 1,\mu_F^2,\lambda) = 
H^{\mathrm{soft}}(z,\mu_F^2,\lambda),
\label{H-Hsoft}
\end{equation}
where
\begin{eqnarray}
H^{\mathrm{soft}}(z,\mu_F^2,\lambda) &=& 2C_F\left\{ 2\left[{\ln(1-z) \over
1-z}\right]_+ - \left[{\ln(1-\lambda z)\over 1-z}\right]_+
\right.\nonumber\\
&+& \left. {1\over 4} \left[{1-z\over (1-\lambda z)^2} \right]_+ +
{1\over (1-z)_+}\left(\ln{Q^2+m^2\over \mu_F^2}-1\right)
\right\},\label{Hsoft}
\end{eqnarray}
with $C_F=4/3$.

One immediately sees that the behaviour of the coefficient function
(\ref{Hsoft}) at large $z$ strongly depends on the value of the
ratio $m/Q$. To illustrate
this point, we consider the two extreme regimes: small-$m/Q$ 
limit, i.e. $\lambda\approx 1$, and large-$m/Q$ limit, i.e.
$\lambda\ll 1$. 

In the first case, setting $\lambda=1$ in
(\ref{Hsoft}) and performing its Mellin transform, 
one recovers the large-$N$ limit of the 
$N$-space $\overline{\mathrm{MS}}$ massless coefficient function $C_N$
reported in \cite{cmw}:
\begin{equation}
C_N^{\mathrm{soft}}(\mu^2,\mu_F^2,Q^2)\vert_{\lambda=1} 
= 1 + {{\alpha_S(\mu^2)C_F}\over{\pi}}
\left\{{1\over 2}\ln^2 N + \left[\gamma_E + {3\over 4}
-\ln{Q^2\over\mu_F^2} \right]\ln N\right\}, \label{softnomass}
\end{equation}
where $\gamma_E=0.577\dots$ is the Euler constant.

In the second case, for $\lambda\to 0$ and $z\to 1$,
one has $\lambda z=\lambda$ in
(\ref{Hsoft}). As a result, the term $\sim [(1-z)/(1-\lambda z)^2]_+$
is regular in the soft limit. The following expression holds in moment space:
\begin{equation}
C^{\mathrm{soft}}_N(\mu^2,\mu_F^2,Q^2,m^2)\vert_{\lambda\ll 1} = 1+
{{\alpha_S(\mu^2)C_F}\over{\pi}} \left\{\ln^2 N+\left[2\gamma_E+1
-\ln{\mathcal{M}^2\over\mu_F^2}\right]\ln N\right\}. \label{dismass}
\end{equation}
For later convenience, we have introduced the scale:
\begin{equation}
\mathcal{M}^2= m^2\left(1+{Q^2\over m^2}\right)^2 \label{M}.
\end{equation}
Furthermore, in evaluating the Mellin transforms, we have
made use of the following relation \cite{ct}:
\begin{eqnarray}
\int_0^1dz z^{N-1}\left[ {\ln^k(1-z)\over (1-z)}\right]_+ &=&
\left\lbrace
 \begin{array}{l l}
  -\ln N + \mathcal{O}(1)&{\rm for}~ k=0, \\
  1/2\ln^2N+\gamma_E\ln N + \mathcal{O}(1) &{\rm for}~ k=1.\nonumber
 \end{array}
\right.
\end{eqnarray}
The different behaviour in the soft limit of the NLO
$\overline{\mathrm{MS}}$ coefficient function for small
and large values of the ratio $m/Q$, which can be seen by comparing
Eqs.(\ref{softnomass}) and (\ref{dismass}),
will lead to different
expressions of the resummed coefficient function, 
which will depend on the value of $m/Q$.
This will be discussed in detail in the next section.

\section{Soft-gluon resummation}

In this section we would like to discuss the resummation of soft-gluon
radiation in the quark-initiated
$\overline{\mathrm{MS}}$ coefficient function of CC DIS.

Soft resummation in DIS can be performed along the general lines
of \cite{ct}: one evaluates the amplitude of the process 
$q_1(p_1)W^*(q)\to q_2(p_2)g(p_g)$ in the eikonal approximation and 
exponentiates the result.

However, we can think of a simpler solution to get the resummed
coefficient function in massive CC DIS.
We note that, in the eikonal approximation, the coefficient function for
heavy quark production in charged-current DIS is related
by crossing to the coefficient function of heavy-quark decay into a $W$ and
a light quark, for example top decay $t(p_t) \to W(q)b(p_b)(g(p_g))$.
In fact, in top decay
the $b$ quark can be treated in the massless approximation 
$m_b\simeq 0$, since $m_b^2\ll m_t^2$ and $m_b^2\ll m_W^2$.

If in top quark decay one observes the $b$ quark, i.e. computes
the differential width with respect to the $b$-quark energy fraction
$x_b$ \cite{cm,ccm}, one recovers a situation which is analogous to CC DIS.
We detect a light quark, namely the $b$ in the top decay coefficient
function and $q_1$ in CC DIS, 
and are inclusive with respect to the heavy quark, the top quark in 
$t$-decay, the massive quark $q_2$ in process (\ref{qs}).

The correspondence with the top-decay process implies that, for large $N$,
the CC DIS large-$m/Q$
coefficient function (\ref{dismass}) coincides with 
the top decay coefficient function 
computed in Refs.~\cite{cm,ccm} after one performs the replacements:
\begin{equation}
m_t\to m\ \ ; \ \ m_W^2\to -Q^2.
\label{mtw}
\end{equation}
We can therefore apply Eq.~(\ref{mtw}), along with
$p_t\to p_2$ and $p_b\to p_1$ in the eikonal current of Ref.~\cite{ccm}
and repeat all the steps which have led to the resummation 
of the top decay $\overline{\mathrm{MS}}$ 
coefficient function,
which we do not report here for the sake of brevity, and
resum soft contributions in CC DIS to next-to-leading logarithmic
(NLL) accuracy. 

As in \cite{ccm}, we write the resummed coefficient function as
an integral over $z$ and $k^2=(p_1+p_g)^2(1-z)$,
with $\mu_F^2\leq k^2\leq  \mathcal{M}^2 (1-z)^2$, and obtain the
following expression:
\begin{eqnarray}
\ln \Delta_N &=& \int_0^1 {dz {{z^{N-1}-1}\over{1-z}}}
\left\{\int_{\mu_F^2}^{ \mathcal{M}^2 (1-z)^2} {{dk^2}\over {k^2}}
A\left[\alpha_S(k^2)\right] +
S\left[\alpha_S\left({\cal M}^2(1-z)^2\right)\right]\right\},
\label{deltares}
\end{eqnarray}
where ${\cal M}$ has been introduced in Eq.~(\ref{M}).

We point out that the upper limit of the integration over $k^2$ can be 
explicitly obtained, e.g.,
in the centre-of-mass frame, where the light-quark energy satisfies 
the relation
\begin{equation}
2E_1\simeq {{Q^2+m^2}\over{\sqrt{(1-z)Q^2+m^2}}}.
\label{upper}
\end{equation}
For finite $m$, in the soft limit $z\to 1$, 
$2E_1\simeq (Q^2+m^2)/m$.
From $k^2\leq 4E_1^2(1-z)^2$, one gets the upper limit 
$k^2\leq {\cal M}(1-z)^2$.

The function $A(\alpha_S)$ in Eq.~(\ref{deltares}) 
can be expanded as a series in $\alpha_S$ as:
\begin{equation}
A(\alpha_S)=\sum_{n=1}^{\infty}\left({{\alpha_S}\over
{\pi}}\right)^n A^{(n)}.
\end{equation}
The first two coefficients are mandatory in order 
to resum our coefficient function
to NLL accuracy \cite{ct}: 
\begin{equation}
A^{(1)}=C_F,
\end{equation}
\begin{equation}
A^{(2)}= {1\over 2} C_F \left[ C_A\left(
{{67}\over{18}}-{{\pi^2}\over 6}\right) -{5\over 9}n_f\right],
\end{equation}
where $C_F = 4/3$, $C_A = 3$ and $n_f$ is the number of quark
flavours. A numerical estimate for $A^{(3)}$ is also known \cite{nv}
(see also \cite{vogt}).

The function $S(\alpha_S)$ is characteristic of processes where a heavy
quark is involved and expresses soft radiation which is not
collinear enhanced.

The following expansion holds:
\begin{equation}
S(\alpha_S)=\sum_{n=1}^{\infty}\left({{\alpha_S}\over
{\pi}}\right)^n S^{(n)},
\end{equation}
where only the first coefficient is needed to NLL level.
It reads:
\begin{equation}
S^{(1)}=-C_F.
\end{equation}
The integral over $z$ in Eq.~(\ref{deltares}) can be performed, up to NLL
accuracy, by making the following replacement \cite{ct}:
\begin{equation}
z^{N-1}-1\to -\Theta\left( 1-{{e^{-\gamma_E}}\over N}-z\right) ,
\end{equation}
where $\Theta$ is the Heaviside step function. 
This leads to writing the following result for the function
$\Delta_N$:
\begin{equation}
\Delta_N(\mu^2,\mu_F^2,m^2,Q^2)=\exp\left[\ln N
g^{(1)}(\ell)+ g^{(2)}(\ell,\mu^2,\mu_F^2)\right]\, ,
\label{deltaint}
\end{equation}
with
\begin{equation}
\ell=b_0\alpha_S(\mu^2)\ln N \, ,
\label{ell}
\end{equation}
and the functions $g^{(1)}$ and $g^{(2)}$ given by
\begin{eqnarray}
g^{(1)}(\ell)&=& \frac{A^{(1)}}{2\pi b_0 \ell} \;
[ 2\ell + (1-2\ell) \ln (1-2\ell)] \;,\label{g1}\\
g^{(2)}(\ell,\mu^2,\mu_F^2) &=& \frac{A^{(1)}}{2 \pi b_0}\left[\ln
\frac{\mathcal{M}^2}{\mu_F^2}
- 2\gamma_E\right] \ln(1-2\ell)\nonumber\\
&+&\frac{A^{(1)}  b_1}{4 \pi b_0^3} \left[ 4\ell + 2 \ln
(1-2\ell) +
\ln^2 (1-2\ell) \right]\nonumber\\
&-& \frac{1}{2\pi b_0} \left[2\ell + \ln (1-2\ell) \right]
\left(\frac{A^{(2)}}{\pi b_0} +
A^{(1)}\ln\frac{\mu^2}{\mu_{F}^2}\right)
\nonumber\\
&+& \frac{S^{(1)}}{2\pi b_0} \ln (1-2\ell).
\label{g2}
\end{eqnarray}
In Eqs.~(\ref{ell}-\ref{g2}), 
$b_0$ and $b_1$ are the first two coefficients of the QCD
$\beta$ function:
\begin{equation}
b_0={{33-n_f}\over{12\pi}},\ \ b_1={{153-19n_f}\over{24\pi^2}}.
\end{equation}
In Eq.~(\ref{deltaint}) the term $\ln N g^{(1)}(\ell)$ accounts
for the resummation of leading logarithms (LL) $\alpha_S^n\ln^{n+1}N$
in the Sudakov exponent, while function
$g^{(2)}(\ell,\mu^2,\mu_F^2)$ resums NLL terms $\alpha_S^n\ln^nN$.
One can also check that the ${\cal O}(\alpha_S)$ expansion of
Eq.(\ref{deltaint}) reproduces the large-$N$ coefficient function in
Eq.~(\ref{dismass}).

Furthermore, we follow \cite{ccm,cc} and also 
include in our final Sudakov-resummed 
coefficient function terms which, in the $N$-space coefficient
function, are constant with respect to $N$ and which we denote 
with $K_i(\mu_F^2,m^2,Q^2)$:
\begin{eqnarray}
H_{N,i}^S(\mu^2,\mu_F^2,m^2,Q^2)&=&
\left[ 1+{{\alpha_S(\mu^2)C_F}\over{2\pi}} K_i(\mu_F^2,m^2,Q^2)\right]
\nonumber\\
&\times& \exp\left[\ln N
g^{(1)}(\ell)+ g^{(2)}(\ell,\mu,\mu_F)\right].
\label{sud}
\end{eqnarray}
The explicit expression for the constant terms reads:
\begin{eqnarray}
K_i(\mu_F^2,m^2,Q^2) &=& \left( {3\over 2}
-2\gamma_E\right)\ln\left({Q^2+m^2 \over \mu_F^2}\right)
+\ln(1-\lambda)\left(
2\gamma_E+{3\lambda-2\over 2\lambda} \right)\nonumber\\
&+& A_i -2{\rm Li}_2\left(-{Q^2\over m^2}\right)
+2(\gamma_E-1)(\gamma_E+2)\label{Kfact},
\end{eqnarray}
where $\lambda$ was introduced in Eq.~(\ref{lamb}) and
we have defined the quantities $A_i$ \cite{gkr}, which are given by:
\begin{equation}
A_1=A_3=0, 
\end{equation}
\begin{equation}
A_2={1\over\lambda} (1-\lambda)\ln(1-\lambda).
\end{equation}
We now match the resummed coefficient function to the exact
first-order result: we add the resummed result to the exact coefficient
function and, in order to avoid double counting, we subtract what
they have in common. Our final result for the resummed coefficient
function reads:
\begin{eqnarray}
H_{N,i}^{\mathrm{res}}(\mu^2,\mu_F^2,m^2,Q^2)
&=&H_{N,i}^S(\mu^2,\mu_F^2,m^2,Q^2)\nonumber\\
&-& \left[H_{N,i}(\mu^2,\mu_F^2,m^2,Q^2)
\right]_{\alpha_S}\nonumber\\
&+& C_{N,i}^q(\mu^2,\mu_F^2,m^2,Q^2)
\label{match}
\end{eqnarray}
where $C_{N,i}^q$ is the Mellin transform of the exact ${\cal O}(\alpha_S)$
coefficient functions $C_i^q$ in Eq.~(\ref{cq}).

Before closing this section, we would like to compare our 
calculation with the results on soft resummation in DIS processes
when the final-state quark $q_2$ is light
\cite{ct,cmw,vogt}. 
In fact, while for top decay, up to width effects, the
value of $m_t^2/m_W^2$ is fixed and rather large, $m^2/Q^2$ can
vary widely in CC DIS according to the value of $Q^2$ in 
the considered process. This implies that, if $m^2\ll Q^2$, the 
heavy quark behaves in a massless fashion and one may have
to use a different result for the resummed coefficient function.

As we have already remarked in Section 2, one cannot just recover the
massless limit from the massive result: the limits $m/Q\to 0$ or
$\lambda\to 1$ of Eqs.~(\ref{deltares}) and (\ref{sud})
appear in fact to be divergent.
The actual result is instead finite: one cannot naively apply 
those equations for very small values of $m/Q$, but rather needs 
to perform a different calculation in this limit.

In fact, if the final-state $q_2$ is massless, the NLL resummed
coefficient function
receives an additional contribution due to purely collinear radiation.  
As shown in \cite{ct}, one can account for such a contribution by solving
a modified evolution equation for the parton distribution
associated with the final state. 

Therefore, in the massless approximation,
one will have to add an additional term to $S(\alpha_S)$ to account
for collinear-enhanced radiation, which leads to the introduction
in the Sudakov exponent
of a function that is usually denoted by $1/2 B[\alpha_S(Q^2(1-z))]$.
The argument of function $B(\alpha_S)$ 
is indeed the virtuality of the final-state
jet, which, for a massless $q_2$ in process (\ref{qs}), will vanish in
soft limit like $(p_2+p_g)^2\simeq (1-z)Q^2$.

On the contrary, if 
the final-state quark is massive, the virtuality of the final state 
in the large-$z$ limit is given by
$(p_2+p_g)^2=(1-z)(Q^2+m^2)+m^2$ and is never small, provided $m/Q$ is 
sufficiently large. We do not have therefore additional collinear 
enhancement for heavy quark production, as we do not have any 
$\sim B(\alpha_S)$ 
contribution in the resummed top decay coefficient function of Ref.~\cite{ccm}.

Since in the following section we shall present results in the limit
$m/Q\to 0$ as well, we explicitly write the resummed coefficient function in
the massless approximation \cite{cmw}.
As for the large-$m/Q$ case, we write it as an integral over $z$ and 
$k^2=(p_1+p_g)^2(1-z)$, with $\mu_F^2\leq k^2\leq Q^2(1-z)$:
\begin{equation}
\ln \Delta_N\vert_{m/Q\to 0} = \int_0^1 {dz {{z^{N-1}-1}\over{1-z}}}
\left\{\int_{\mu_F^2}^{Q^2 (1-z)} {{dk^2}\over {k^2}}
A\left[\alpha_S(k^2)\right] +
{1\over 2} 
B\left[\alpha_S\left(Q^2(1-z)\right)\right]\right\}.
\label{deltazero}
\end{equation}
The upper limit for the integration variable $k^2$ 
can be obtained using Eq.~(\ref{upper}): for $m=0$ and
$z\to 1$, one has $2E_1\simeq Q/\sqrt{1-z}$.
From $k^2\leq 4E_1^2(1-z)^2$, the relation $k^2\leq Q^2(1-z)$ follows.

We can expand function $B(\alpha_S)$ as a series in $\alpha_S$: 
\begin{equation}
B(\alpha_S)=\sum_{n=1}^{\infty}\left({{\alpha_S}\over
{\pi}}\right)^n B^{(n)}
\end{equation}
and, to NLL level, keep only the first term of the expansion \cite{ct}:
\begin{equation}
B^{(1)}=-{3\over 2}C_F.
\end{equation} 
Explicit results for Eq.~(\ref{deltazero}) to NLL level can be found
in Ref.~\cite{vogt}, where functions analogous to our $g^{(1)}$ and 
$g^{(2)}$ are reported \footnote{We point out that Ref.~\cite{vogt}
contains results to next-to-next-to-leading logarithmic (NNLL) accuracy
for soft resummation in massless DIS. However, 
for the sake of comparison with the
massive result (\ref{deltares}), we shall use in the following 
the results of \cite{vogt} to NLL.}.
One can also check that the ${\cal O}(\alpha_S)$ expansion of 
Eq.~(\ref{deltazero})
yields the large-$N$ limit of the NLO coefficient function presented in 
Eq.~(\ref{softnomass}). 

As in Eq.~(\ref{sud}), we also include in the final Sudakov-resummed 
massless coefficient function terms which are constant with respect 
to $N$. Once again, we observe that such constant terms cannot be obtained 
just taking the $m/Q\to 0$ limit of the massive case, i.e. 
Eq.~(\ref{Kfact}), but we shall have 
to explicitly extract them from the massless NLO coefficient function.
We obtain:
\begin{equation}
K_i(\mu_F^2,m^2,Q^2)\vert_{m/Q\to 0} = \left( {3\over 2}
-2\gamma_E\right)\ln\left({Q^2 \over \mu_F^2}\right) + \gamma_E^2 +
{3\over 2}\gamma_E - {\pi^2\over 6} - {9\over 2},
\label{Kfactor-zero}
\end{equation}
for $i=1,\ 2,\ 3$.
As in Eq.~(\ref{match}) we  match 
the resummed result (\ref{deltazero}) to the exact coefficient
function (\ref{cq}), with functions $H_i^q$ now evaluated in the limit
$m/Q\to 0$, i.e. $\lambda\to 1$.

More details on the derivation of Eq.~(\ref{deltares}) as well as on the
comparison with the massless result (\ref{deltazero}) can be found in
\cite{mitov}.

\section{Phenomenological results}
We would like to present results on proton structure functions for charm quark 
production in CC DIS and investigate the effects of
soft-gluon resummation. 
As discussed in Section 2, structure functions are given by convolutions
of $\overline{\mathrm{MS}}$ coefficient functions with parton distribution
functions.
Since the resummed coefficient function is given in $N$-space, in principle
one may like to use parton distribution functions in Mellin space as well,
in order to get the resummed structure function in moment space 
and finally invert it numerically from $N$- to $x$-space.
However, all modern sets of parton distribution functions 
\cite{cteq,mrst,grv} are given numerically
in the form of grids in the $(x,Q^2)$ space.

In order to overcome this problem, we follow the method proposed in
Ref.~\cite{csw} in the context of joint resummation, which allows
one to use $x$-space parton distributions, even when performing
resummed calculations in Mellin space.

Such a method consists of rewriting the integral of the
inverse Mellin transform of the resummed structure
functions as follows:
\begin{eqnarray}
{\cal F}_i^{\mathrm{res}}(x,Q^2)&=&
{1\over{2\pi i}}\int_{\Gamma_N}{dN \chi^{-N} q_N(\mu_F^2)
H_{N,i}^{\mathrm{res}}(\mu^2,\mu_F^2,m^2,Q^2)}\nonumber\\
&=&
{1\over{2\pi i}}\int_{\Gamma_N}{dN \chi^{-N} 
[(N-1)^2q_N(\mu_F^2)]
{{H_{N,i}^{\mathrm{res}}(\mu^2,\mu_F^2,m^2,Q^2)}\over{(N-1)^2}}},
\end{eqnarray}
where $q_N(\mu_F^2)$ is the parton distribution of the initial-state quark
$q_1$, 
as if it were available in $N$ space, and $\Gamma_N$ is the integration
contour in the
complex plane, chosen according to the Minimal Prescription
\cite{cmnt}.

One can integrate out the term with the parton distribution function by
observing that the following relation holds:
\begin{equation}
{1\over{2\pi i}}\int_{\Gamma_N}{dN \xi^{-N} (N-1)^2q_N (\mu_F^2)}=
{d\over{d\xi}}\left\{\xi {d\over{d\xi}}\left[\xi q(\xi,\mu_F^2)\right]\right\}=
\Phi(\xi,\mu_F^2).
\label{phi}\end{equation}
Due to the above relation, one does not need anymore the $N$-space
parton distributions, but one should just differentiate the $x$-space ones,
which numerically can be done. 

As one has the analytical expression for $H_{N,i}^{\mathrm{res}}$
in $N$-space (\ref{match}), the following inverse Mellin transform is
straightforward:
\begin{equation}
{\cal H}(\xi,\mu_F^2)={1\over{2\pi i}}\int_{\Gamma_N}
{dN\xi^{-N}{{H_{N,i}^{\mathrm{res}}(\mu^2,\mu_F^2,m^2,Q^2)}\over{(N-1)^2}}}
\label{calh}
\end{equation}
which, thanks to the suppressing factor $1/(N-1)^2$ in the integrand,
turns out to be smooth for $\xi\to 1$.

The resummed 
structure function will be finally expressed as the following convolution:
\begin{equation}
{\cal F}_i^{\mathrm{res}}(x,Q^2)=
\int_\chi^1{{d\xi}\over{\xi}}{\cal H}(\xi,\mu^2,\mu_F^2)\Phi
\left({\chi\over\xi},\mu_F^2\right).
\end{equation} 

After having clarified how to deal with the parton distribution functions,
we are able to present results for soft-resummed structure functions.
We shall consider charm quark production, i.e. $q_2=c$ in Eqs.~(\ref{born})
and (\ref{qs}), since processes with charm quarks in the final state 
play a role for structure function measurements
at NuTeV and HERA experiments. 

As for the choice of the parton distribution set, in principle, once
data on heavy quark production in CC DIS is available, one 
should use the NLL coefficient functions when performing the
parton distribution global fits and get NLL parton densities as well.
For the time being, however, we can just use
one of the most-updated NLO sets and convolute it with
the fixed-order or resummed coefficient function. 

We shall present results based on the
new generation of CTEQ NLO $\overline{\mathrm{MS}}$
parton distribution functions \cite{cteq}, the so-called
CTEQ6M set, but similar
results can be obtained using, e.g., the MRST \cite{mrst} or the GRV 
\cite{grv} sets.

The elementary scattering processes which yield the production of charm
quarks in CC DIS are $dW^*\to c$ and $sW^*\to c$.
For $e^+(e^-)p$ scattering at HERA, our parton distribution function  
$q_1(\xi,Q^2)$ in Eq.~(\ref{conv}) will be as follows:
\begin{equation}
q_1(\xi,Q^2)\vert_{\mathrm{HERA}}=|V_{cd}|^2d(\xi,Q^2)+|V_{cs}|^2s(\xi,Q^2),
\label{q1}
\end{equation}
where $V_{cd}$ and $V_{cs}$ are the relevant 
Cabibbo--Kobayashi--Maskawa matrix elements.
For neutrino scattering on an isoscalar target, Eq.~(\ref{q1}) will have to
be modified in order to account for the possibility of an interaction
with a neutron as well:
\begin{equation}
q_1(\xi,Q^2)\vert_{\mathrm{NuTeV}}=|V_{cd}|^2
{{d(\xi,Q^2)+u(\xi,Q^2)}\over 2}+|V_{cs}|^2s(\xi,Q^2),
\end{equation}
where $d(\xi,Q^2)$, $u(\xi,Q^2)$ and $s(\xi,Q^2)$ are still the proton 
parton distribution functions and we have applied isospin symmetry, i.e. 
$u_p=d_n$ and $s_p=s_n$. 

In order to be consistent with the use of the CTEQ parton distribution
functions, we shall use for the QCD scale in the
$\overline{\mathrm{MS}}$ renormalization scheme
the values $\Lambda_4=326$~MeV and
$\Lambda_5=226$~MeV, for four and five active flavours respectively.
This corresponds to $\alpha_S(m_Z)\simeq 0.118$.
The charm and bottom quark masses have been set to 
$m_c=1.3$~GeV and $m_b=4.5$~GeV, as was done in \cite{cteq}.

Since mass effects are important for large values of
the ratio $m_c^2/Q^2$, we would like to have $Q$ as small as possible 
so as to be able to apply our massive resummation.
The NuTeV experiment is indeed able to measure structure functions 
at small $Q^2$ values \cite{nutev,nutev1}, as it can detect the
final-state muon produced in $\nu_\mu N\to \mu X$ processes.
In our phenomenological analysis we shall consider charm production
at $Q^2=2$~GeV$^2$ and $Q^2=5$~GeV$^2$, which are values
reachable at NuTeV such that $m_c^2/Q^2$ is relatively large.

The detection of CC events at HERA is more problematic,  
due to backgrounds and to the presence of a 
neutrino, instead of a charged lepton, in the final state.
The current high-luminosity HERA II experiments may detect 
heavy quarks in CC events, but the $Q^2$ values for such events
are still supposed to be much larger than the charm quark mass squared,
typically $Q^2\gsim 100$~GeV$^2$ \cite{ck}.
We shall nonetheless present results for charm quark production at HERA,
but in this case we shall have to use the massless result for the
resummed $\overline{\mathrm{MS}}$ coefficient function, i.e. 
Eq.~(\ref{deltazero}) and the formulas reported in \cite{vogt}.
We shall consider the typical HERA values
$Q^2=300$~GeV$^2$ and 1000 GeV$^2$.

We present results for the structure function 
$F_2^c(x,Q^2)$, but we can already anticipate that the effect of the 
resummation is approximately the same on all three structure functions
$F_1^c$, $F_2^c$ and $F_3^c$ and on the single-differential cross
section $d\sigma/dx$, obtained from Eq.(\ref{sigma}), after integrating over 
$y$.

For the sake of comparison with the experiments, we shall plot 
the structure function $F_2^c$ in terms of the Bjorken $x$, which 
is the measured quantity. For most of our plots,  
we shall set renormalization and factorization scales equal to $Q$, i.e.
$\mu_F=\mu=Q$. Afterwards, we shall also investigate 
the dependence of our results on the choice of such scales.
\begin{figure}
\centerline{\resizebox{0.65\textwidth}{!}{\includegraphics{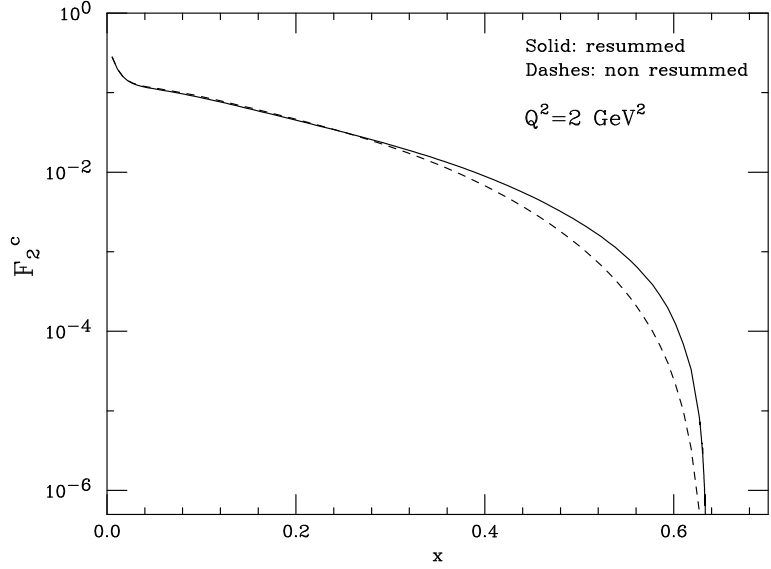}}}
\caption{\small Results for $F_2^c(x,Q^2)$ for charm quark
production in neutrino scattering at $Q^2=2$~GeV$^2$ 
with (solid) and without (dashed) soft resummation in the
coefficient function. We have set $\mu_F=\mu=Q$.}
\label{q2}
\end{figure}
\begin{figure}
\centerline{\resizebox{0.65\textwidth}{!}{\includegraphics{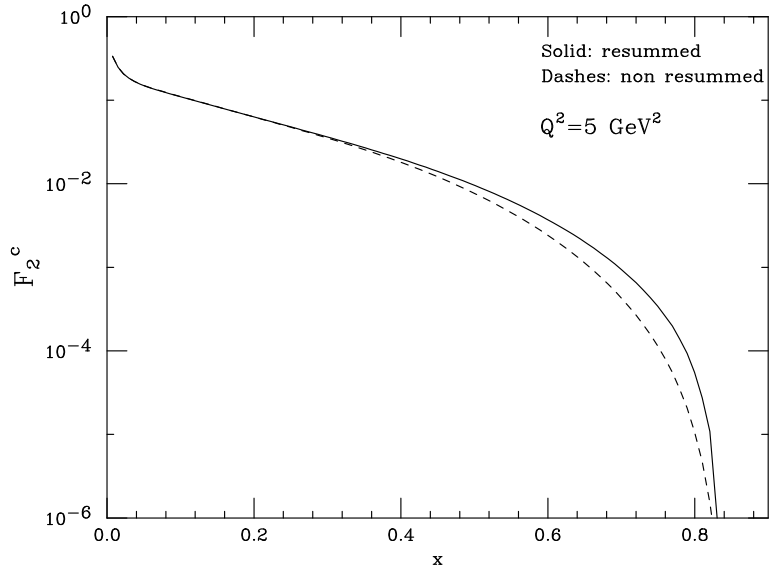}}}
\caption{\small As in Fig.~\ref{q2}, but
for $Q^2=5$~GeV$^2$.}
\label{q5}
\end{figure}

In Figs.~\ref{q2} and \ref{q5} we show $F_2^c(x,Q^2)$ 
in neutrino scattering at $Q^2=2$~GeV$^2$ and 5 GeV$^2$
and compare the results obtained using fixed-order and 
soft-resummed $\overline{\mathrm{MS}}$ coefficient functions. 
We have 
set the target mass to $M=1$~GeV, which is a characteristic nucleon mass.
We observe a relevant effect of the implementation of
soft resummation: the two predictions agree up to $x\simeq 0.2-0.3$, 
afterwards one can see an enhancement of the structure function due
to the resummation. 
At very large $x$, resummed effects are indeed remarkable: for $Q^2=2$~GeV$^2$ 
one has an enhancement of a factor of 2 at $x=0.5$ and a factor of 5 at 
$x=0.6$. For $Q^2=5$~GeV$^2$ one gets a factor of 2 at $x=0.7$ and 5
at $x=0.8$.

\begin{figure}
\centerline{\resizebox{0.65\textwidth}{!}{\includegraphics{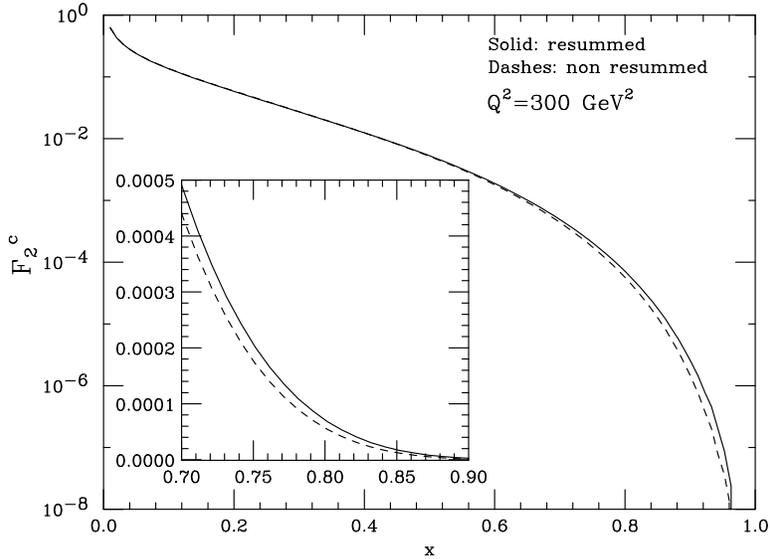}}}
\caption{\small Results for $F_2^c(x,Q^2)$ for positron-proton scattering
at $Q^2=300$~GeV$^2$, with (solid) and without (dashes) 
soft resummation in the coefficient function.
We have set $\mu_F=\mu=Q$.
In the inset figure, we show the same plots at large
$x$ and on a linear scale.}
\label{q300}
\end{figure}
\begin{figure}
\centerline{\resizebox{0.65\textwidth}{!}{\includegraphics{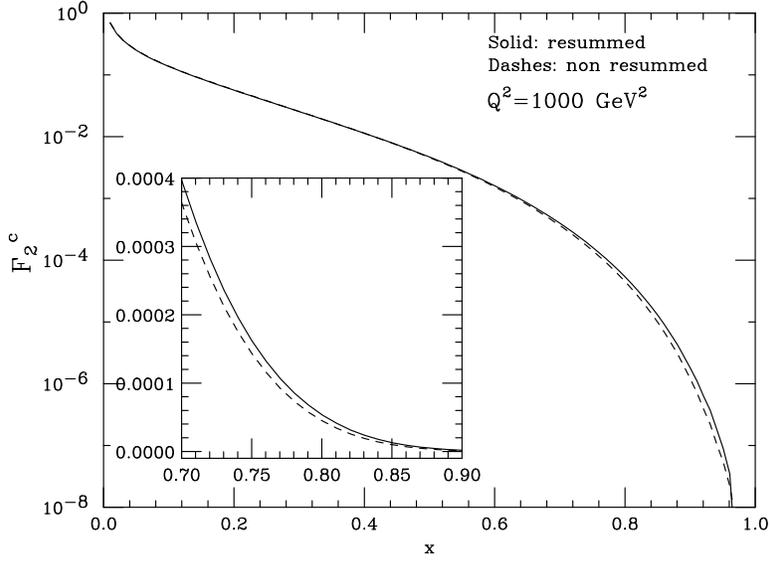}}}
\caption{\small As in Fig.~\ref{q300}, but
for $Q^2=1000$~GeV$^2$.}
\label{q1000}
\end{figure}

In Figs.~\ref{q300} and \ref{q1000} 
we present results for $F_2^c(x,Q^2)$, but for charm 
production at HERA, in particular for positron-proton scattering
at $Q^2=300$ and 1000 GeV$^2$.
In this case, since $m_c/Q\ll 1$, we use the massless result (\ref{deltazero}) 
for the resummed coefficient function.
We see that the impact of the resummation is 
smaller than in the case of low $Q^2$ values. This is a reasonable result:
in fact, leading and next-to-leading logarithms in the Sudakov exponent are
weighted by powers of $\alpha_S(\mu^2)$. As, for example, 
$\alpha_S(2~\mathrm{GeV}^2)\simeq 3\  \alpha_S(300~\mathrm{GeV}^2)$,
resummed effects are clearly more important when $Q^2$ is small. Moreover, 
the larger $Q^2$ is, the larger the values of $x$ are
at which one is sensitive to Sudakov effects.

We note in Figs.~(\ref{q300}) and (\ref{q1000}) that,  
for $x>0.6$, fixed-order and resummed predictions 
start to be distinguishable.
We estimate the overall 
impact of large-$x$ 
resummation on $F_2^c(x,Q^2)$ at $Q^2=300$~GeV$^2$ and 
$Q^2=1000$~GeV$^2$ to be between $10\%$ and $20\%$.

\begin{figure}
\centerline{\resizebox{0.65\textwidth}{!}{\includegraphics{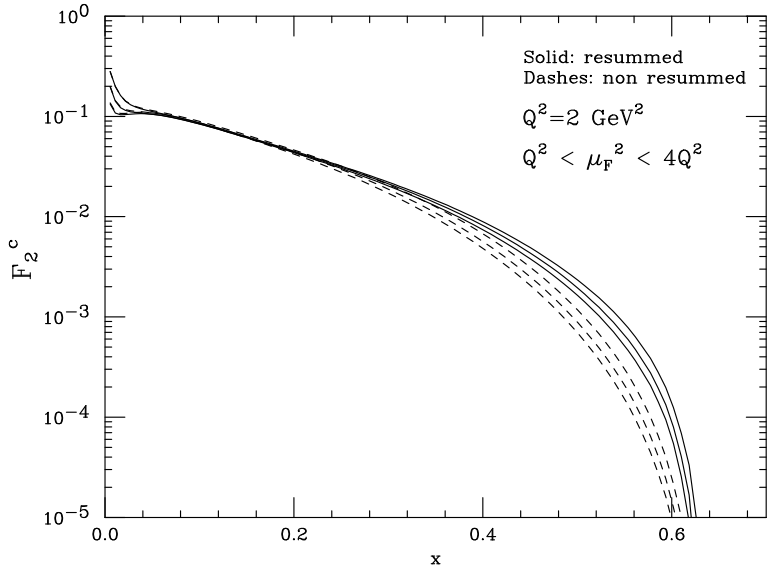}}}
\caption{\small Dependence of $F_2^c$ on the factorization scale for
neutrino scattering at $Q^2=2$~GeV$^2$.
Solid lines include soft resummation in the coefficient function,
dashed lines are fixed-order predictions.}
\label{facnu}
\end{figure}

\begin{figure}
\centerline{\resizebox{0.65\textwidth}{!}{\includegraphics{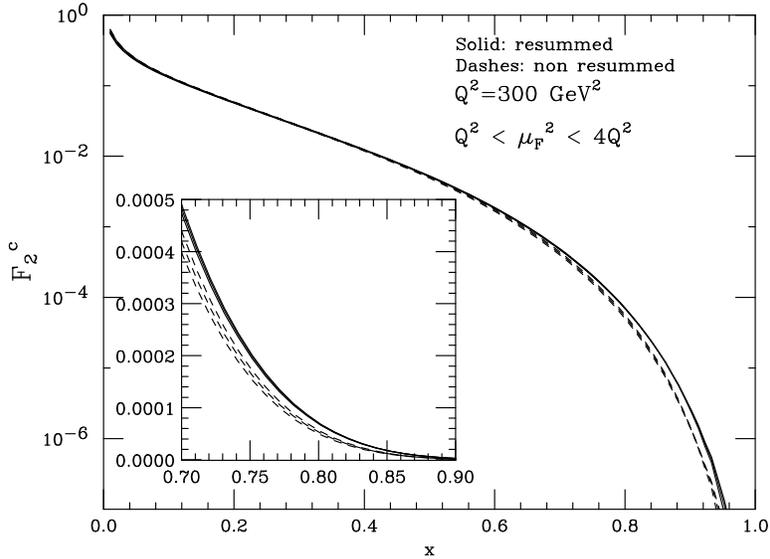}}}
\caption{\small As in Fig.~\ref{facnu}, but for $e^+p$ scattering at
$Q^2=300$~GeV$^2$. In the inset figure, we plot the same curves at large
$x$ and on a linear scale.}
\label{fach}
\end{figure}

In Figs.~\ref{facnu} and \ref{fach} 
we investigate the dependence on factorization and renormalization
scales. We still keep $\mu=\mu_F$, but we allow such scales to assume
the values $Q^2$, $2Q^2$ and $4Q^2$.
We consider $Q^2$ values of 2 and 300 GeV$^2$ for the 
experimental environments of NuTeV and HERA respectively.
We see that the curves which implement soft resummation in the coefficient 
function show a weaker dependence on the chosen value of the 
factorization/renormalization scales. 
In Fig.~\ref{facnu} one can see that one still has a visible effect of 
the value of such scales, but the overall dependence on
$\mu_F$ and $\mu$ of the resummed prediction is smaller than for the 
fixed-order.
The plots at large $Q^2$ exhibit, in general, a weak dependence on
$\mu_F$ and $\mu$ even at NLO, as shown in Fig.~\ref{fach}:
in fact, the dependence on the factorization and renormalization scales is
logarithmic, and hence smaller once $\mu$ and $\mu_F$ vary around large 
values of $Q$.
Nevertheless, while the NLO structure function 
still presents a residual dependence on the scales, the resummed predictions
obtained with three different values of $\mu_F$ are basically
indistinguishable.
A smaller dependence on such scales implies a reduction of
the theoretical uncertainty of the prediction and is
therefore a remarkable
effect of the implementation of soft gluon resummation.

\begin{figure}
\centerline{\resizebox{0.65\textwidth}{!}{\includegraphics{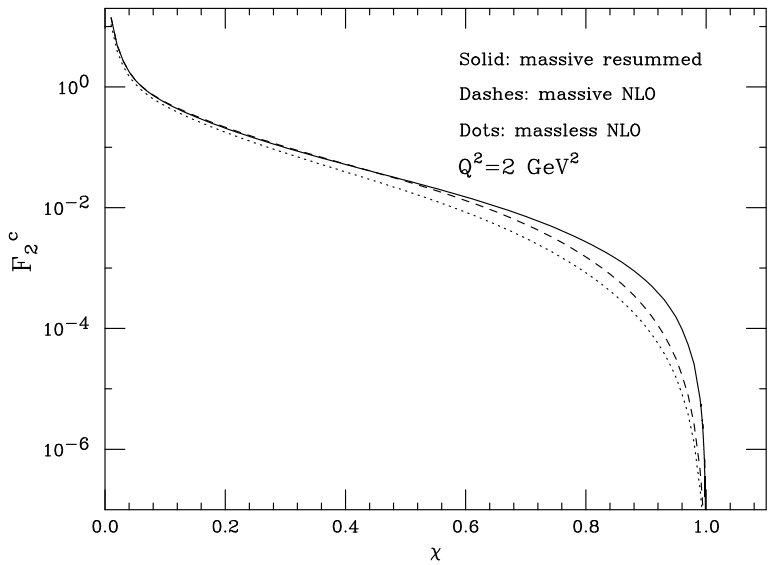}}}
\caption{\small Comparison of massive (dashes) and massless (dots) 
fixed-order calculations with the resummed massive result (solid)
for $Q^2=2$~GeV$^2$ and $\mu=\mu_F=Q$. Plotted is the theoretical
structure function ${\cal F}_2^c(\chi,Q^2)$ defined in Eq.(\ref{conv}).}
\label{mpl}
\end{figure}

We finally would like to compare the impact that
mass effects and soft-gluon resummation have on the charm structure
functions. To achieve this goal, we plot in Fig.~\ref{mpl}
the theoretical structure 
function ${\cal F}_2^c(\chi,Q^2)$ defined in Eq.~(\ref{conv})
as a function of the variable $\chi$ (see Eq.~(\ref{chi})), 
for neutrino scattering at $Q^2=2$~GeV$^2$.
We compare fixed-order
massive (dashed line), fixed-order massless (dots) 
and massive resummed (solid) predictions. 
We observe that the two fixed-order calculations yield different
predictions throughout the full $\chi$ range, which is a consequence of the 
implementation of mass effects; however, at large $\chi$, 
where one starts to be sensitive to the resummation, the impact of
soft resummation is competitive with the one of mass contributions.

Before closing this section, we would like to point out that, although we
have improved our perturbative 
prediction implementing soft-gluon resummation in 
the coefficient function, non-perturbative corrections are still missing.
Non-perturbative effects are important especially at small values
of $Q^2$ and large $x$. 
In fact, a more accurate investigation of the very-large $x$ limit of the
structure functions shows that they exhibit an oscillatory 
behaviour once $x$ gets closer to the maximum value which is kinematically
allowed.

\section{Conclusions}
We have studied inclusive heavy quark production in charged-current Deep 
Inelastic Scattering. We have reviewed the results for the 
NLO cross section and structure functions
and observed that the 
$\overline{\mathrm{MS}}$ quark-initiated coefficient function
contain terms which become
arbitrarily large once the initial-state light-quark energy fraction
approaches unity, which corresponds to soft-gluon radiation.
We have resummed soft-gluon contributions
in the $\overline{\mathrm{MS}}$ coefficient function
to next-to-leading logarithmic accuracy.
We have compared the results
with the resummed coefficient function for $b$-quark production
in top decay and for CC DIS processes, 
but with a light quark in the final state.

We have presented results on the structure function $F_2^c(x,Q^2)$ for
charm quark production at NuTeV and HERA, using the massive or the massless
coefficient function according to the value of the ratio $m_c/Q$. 
We have compared predictions obtained considering
fixed-order and resummed coefficient functions and have found a remarkable
effect due to soft resummation. The structure functions exhibit an enhancement
at large $x$, which is visible especially when $Q$ is comparable with the
heavy quark mass.
The results including soft resummation exhibit very little
dependence on the choice of factorization and renormalization scales, 
which is a reduction of the theoretical uncertainty.
Furthermore, we have compared massive and massless predictions at small $Q^2$
and have found that, at large $x$, resummation effects are of a 
magnitude similar to mass corrections.

It will be clearly very interesting to compare our predictions 
with experimental data. New data on NuTeV structure functions will be soon
available \cite{naples} and we plan to investigate whether the structure
functions yielded by our calculation are able to fit the data.
In particular, we wish to evaluate the role played by soft-gluon
resummation in performing such fits. 
The preliminary analyses in Ref.~\cite{nutev} show in fact
that the NuTeV data lies above the NLO structure functions
based on Ref.~\cite{thorne} for $x\gsim 0.55$.
The effect due to the resummation is indeed an enhancement of structure 
functions at large $x$, hence it may throw some light on the NuTeV
studies. However, for such a comparison to be reliable, we shall have
to implement nuclear-correction effects into our structure functions,
to account for the neutrino scattering on an iron target.
A similar comparison can be made with the foreseen 
CC data from HERA II. 
Moreover, we believe that once such a data becomes
available, our calculation can also be used to perform updated NLL 
global fits of
the parton distribution functions.

\section*{Acknowledgements}
We acknowledge M. Cacciari for providing us with the computer code to
perform inverse Mellin-space transforms and his collaboration in the
early stages of this project.
We are also grateful to A. Bodek, C. Kiesling, A.K. Kulesza, D. Naples,
L.H. Orr and S. Kretzer for discussions on these and related topics.
The work of A.D.M. was supported in part 
by the U.S. Department of Energy, under grant DE-FG02-91ER40685.

\end{document}